\documentclass[11pt,twoside]{article}
\usepackage{asp2004}
\usepackage{graphics}
 \makeindex
\begin{document}
\title{New Systems Showing Light-Time Effect}
\author{P. Zasche}
\affil{Astronomical Institute, Charles University, V
Hole\v{s}ovi\v{c}k\'{a}ch~2, Prague~8, 180 00, Czech Republic}

\begin{abstract}Two Algol-type eclipsing binary systems (EW Lyr and IV Cas) have
been investigated for period changes. Our study was primarily
focused on the light-time effect with alternative explanation by
magnetic activity cycles. In~the case of EW Lyr we have found
third body in the orbit with period about 78 years, amplitude
$A=0.052$ days and orbital eccentricity $e=0.57$. For IV~Cas the
long period is 58 years, amplitude $A=0.034$ days and zero
eccentricity. With these results we are also able to calculate
mass functions and minimal masses of these components.
\end{abstract}

\section{Introduction\label{sec:intro}}
Eclipsing binaries are known as ideal cosmic laboratories for
studying the properties of stars. One method is the period
analysis of these systems. In some cases there is a possibility of
detection of other bodies in these systems using this method. But
the $O\!-\!C$ diagrams have to show periodic oscillations with
well-defined shapes, sufficiently large amplitudes and adequately
long periods. Very short periods have only small amplitudes and
the long periods are not well covered with data yet. All these
conditions and the accuracy of our data set have to be considered
for the limitations of our proposed hypothesis of the light-time
effect. \index{$O-C$ diagram}

The light-time effect (hereafter LITE), caused by the orbital
motion of the eclipsing pair around the barycenter of the multiple
system, was   explained by Irwin (1959) and its necessary
criteria have been listed by Friebos-Conde \& Hertzeg (1973) and
also by Mayer (1990). Presence of a third body in the system is
possible only if the times of minimum, also the secondary ones,
behave in agreement with a theoretical LITE curve, the resultant
mass function has reasonable value and corresponding variations in
radial velocities are measured. During the last decade also a
confirmation by astrometry seems to be plausible. But regrettably
in both investigated systems (EW Lyr and IV Cas) no photometric or
spectroscopic study was performed, so these LITE systems are still
only hypothetic, because of no other confirmation of the LITE. But
there is still a possible explanation of this variations in minima
times by another phenomenon, such as magnetic activity. We discuss
this possibility later. \index{light-time effect} \index{EW Lyr}
\index{IV Cas}

In both cases we have calculated new light elements of the
eclipsing pair and also the parameters of the third body orbit.
Their $O-C$ diagrams are shown in the Figs.~1 and 2. The
curves represent computed light-time effect caused by the third
bodies with parameters given in Table~2 below. The individual
primary and secondary minima are denoted by dots and circles,
respectively. Larger symbols correspond to the photoelectric or
CCD measurements which were given higher weights in our
computations.

For the computation we used the least-squares method. All data
were found in published papers. Table 2 presents our results,
where $M_i$ are masses of components, $p_3$ computed period of the
unresolved body, $A$ semiamplitude of LITE, $e$ eccentricity,
$\omega$ length of periastron, $f(M_3)$ computed mass function and
$M_{3,{\rm min}}$ minimum mass (for $i$ = 90$^\circ$) of the third
body, respectively.

\section{EW Lyr\label{EWLyr}}
EW Lyrae is an Algol-type eclipsing binary discovered by
Hoffmeister (1930) and its period changes were recognized in
Szafranic (1956). Spectral type was determined as F0 (Brancewicz
\& Dworak 1980) or A8--9V (Halbedel 1984) with the individual
masses about 1.39 and 0.84 $\mathrm{M_\odot}$ according to Budding
(2004). But the values are still not very convincing. Our
collection of minimum times (covering more than 70 years) leads to
new recomputed light elements
\[ \mathrm{Min\:I} = \mathrm{HJD}\;\:24\;26499.7224 + 1.\!\!^{\rm
d}94873888 \cdot {\rm E}\] The potential third body orbits the
eclipsing pair with a period of almost 78 years and eccentricity
$e=0.57$. Its predicted minimal mass would be about
1.1~$\mathrm{M_\odot}$. So in the solution of the light curve, the
third light would be measurable, but regrettably no light-curve
study of this system was made yet.

\begin{figure}
 \scalebox{0.75}{\includegraphics*[17mm,100mm][193mm,181mm]{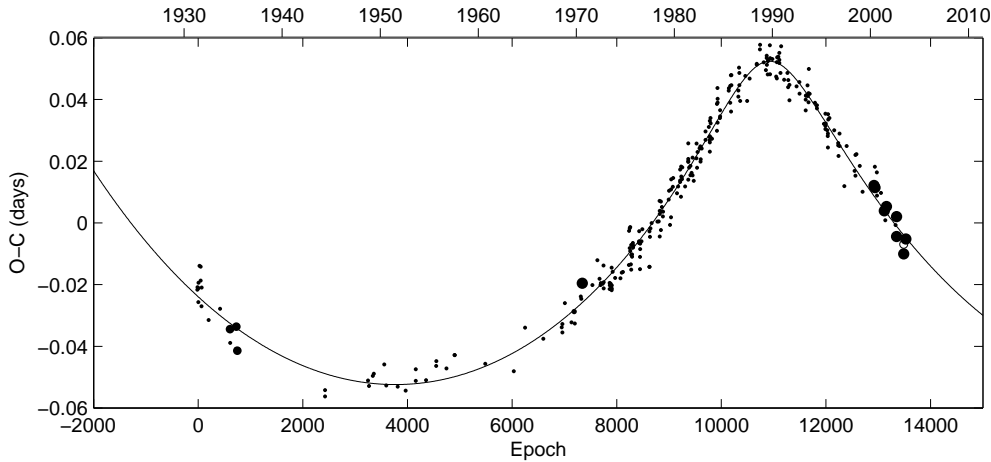}}
 \caption[]{The $O-C$ diagram of EW Lyr.}
 \label{EWLyr}
\end{figure}

We could also discuss so-called Applegate mechanism which causes
the periodic variations in the $O-C$ diagram. At the very
beginning we have to accept some assumptions about the basic
parameters of the system: $M_1=1.39 \mathrm{M_\odot}$, $M_2=0.84
\mathrm{M_\odot}$, $R_1=1.59 \mathrm{R_\odot}$, $R_2=0.85
\mathrm{R_\odot}$. But these values are taken only from
statistical properties of stars of each spectral type, so they are
not very convincing. We also assume the separation of the
components to be $a=10 \mathrm{R_\odot}$ and the rate of the
differential rotation of the stars $\Delta \Omega \simeq
\Omega_{dr}$, see Applegate (1992). With these parameters we are
able to determine the properties of the Applegate effect for both
components (see Table 1). \index{Applegate mechanism}

But as we can see from the values, this explanation seems to be
completely improbable. Especially if we focus on the values
$\Delta L_{\mathrm{RMS}}$ and $B$, they are very high and make no
sense in our case. Luminosity variations caused by this effect
could not exceed the total luminosity of the system and also the
value of magnetic field seems to be quite high. 
 So we can clearly conclude that this effect is not presented in EW
Lyr.

\begin{figure}[b!]
 \scalebox{0.75}{\includegraphics*[17mm,100mm][196mm,181mm]{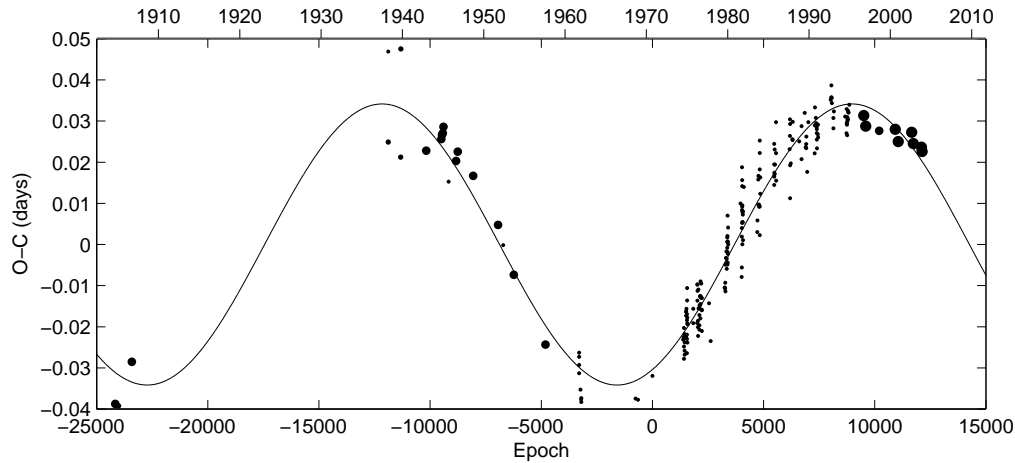}}
 \caption[]{The $O-C$ diagram of IV Cas.}
\label{IVCas}
\end{figure}

\section{IV Cas}
IV Cas is an Algol-type eclipsing binary, with spectrum classified
as A2. As in the case of EW Lyr, our knowledge about the system
parameters is not very satisfactory. From the data we have
calculated new light elements
\[ \mathrm{Min\:I} = \mathrm{HJD}\;\:24\;40854.6289 + 0.\!\!^{\rm
d}99851644 \cdot {\rm E}\] The $O-C$ diagram in Figure 2 covers
one hundred years. Until the 1990s there were a possibility to
describe the shape of the $O-C$ diagram by a quadratic term, so
the mass-transfer hypothesis was discussed. But with new
accurately measured minima we are nowadays sure about the periodic
behavior. We have also neglected the visual minima times measured
after 1995. The potential third body was found in the circular
orbit of period almost 60 years and amplitude 0.034 days. This
gives the value of minimal mass about 1.16$\mathrm{M_\odot}$. But
as in the case of EW Lyr, we do not have any analysis of a light
curve and also any spectroscopic study of this system, so we
cannot prove this result. \index{Algol-type}

With the presumptions  $M_1=2.6 \mathrm{M_\odot}$, $M_2=1.24
\mathrm{M_\odot}$, $R_1=2.0 \mathrm{R_\odot}$, $R_2=2.22
\mathrm{R_\odot}$ (see Budding 2004),  the separation of the
components $a=10 \mathrm{R_\odot}$ and differential rotation of
the stars $\Delta \Omega \simeq \Omega_{dr}$  we are able to
derive the parameters of the Applegate effect (given in Table 1).
\begin{table}[h!]
 \caption[]{The parameters of the Applegate effect for both stars.}
 \label{tabl}
 \scalebox{0.99}{
{\footnotesize
\begin{tabular}{cccccc}
\hline \bfseries $\:$ & $\:$ EW Lyr$\mathrm{_{Prim}}$ $\,$ & $\,$ EW Lyr$\mathrm{_{Sec}}$ $\,\,$ & $\:$ IV Cas$\mathrm{_{Prim}}$ $\,$ & $\,$ IV Cas$\mathrm{_{Sec}}$ $\:$ & $\,$ Units$\,$ \\
\hline
$\Delta P$ & 1.95 & 1.95 & 0.88 & 0.88 & s \\
$I_{\mathrm{eff}}$ & $7.55$ & $7.80$ & $33.4$ & $19.6$ &
$10^{53}{\mathrm{g}\!}\cdot{\!\mathrm{cm}}^2$ \\
$\Delta J$ & $34.6$ & $5.65$ & $14.9$ & $2.48$ &
$10^{48}{\mathrm{g}\!}\cdot{\!\mathrm{cm}^2}\!\cdot{\!\mathrm{s}^{-1}}$\\
$\Delta \Omega$ & $229$ & $36.2$ & $22.3$ & $6.32$ &
$10^{-7}{\mathrm{s}^{-1}}$ \\
$\Delta E$ & $1580$ & $40.9$ & $66.7$ & $3.14$& $10^{42}$erg \\
$\Delta L_{\mathrm{RMS}}$  & 520 & 13.4 & 29.5 & 1.4 & ${\mathrm{L_\odot}}$\\
$B$ & $59.9$ & $16.2$ & $23.9$ & $8.3$ & kG\\ \hline
\end{tabular}}}
\end{table}

But as we can see from the values, this explanation seems to be
again improbable on primary component, but we could discuss, if
affects the secondary or not. For example the value $\Delta
L_{\mathrm{RMS}}$ seems to be rather high, but if we compare it
with the luminosity of the secondary, we almost satisfy  the
condition $\Delta L_{\mathrm{RMS}} / L \la 0.1$ introduced by
Applegate (1992), because we get $\Delta L_{\mathrm{RMS}} \simeq
0.18 L$. So we could discuss the probability of period modulation
in IV Cas caused by the magnetic activity cycles on secondary
component. This possibility remains to be studied and confirmed to
the future.

\begin{table}[h!]
 \caption[]{The resulting LITE parameters for investigated systems.}
 \label{tabl}
 \scalebox{0.83}{
 \begin{tabular}{cccccccccc}
 \hline
Name of star & Spectrum &  $ M_1$     &  $ M_2$     & $ p_3 $ &  $A_3$  & $e_3$ &$\omega_3$&  $f(M_3) $  & $M_{3,{\rm min}}$ \\
             &          & [$M_\odot$] & [$M_\odot$] & [years] &  [days] &       &  [deg]   & [$M_\odot$] & [$M_\odot$] \\
\hline
 EW Lyr     &    A8-9V  &    1.39     &     0.84    &  77.83  &  0.0524 & 0.574 &   86.4   &   0.123     &     1.11    \\
 IV Cas     &     A2    &    2.60     &     1.24    &  57.78  &  0.0342 & 0.001 &    5.8   &   0.062     &     1.16    \\
\hline
\end{tabular}}
\end{table}

\section{Conclusions}
We have derived new LITE parameters for two eclipsing binaries by
means of the $O-C$ diagram analysis. We have suggested the
light-time effect for explanation of the period variation. We have
also mentioned a probability of presence of magnetic activity
cycles, which could modulate the period. In case of secondary
component of IV Cas this hypothesis should also be in stage, but
we have not enough data to make a final decision. So the
consequence is, that for the confirmation of presence of the LITE
in these systems, we need further detailed photometric,
spectroscopic or astrometric study of these binaries.


{}

\end{document}